%% file: geach_sep08_astroph.tex
\documentclass{emulateapj}
\usepackage{psfig}
\usepackage{apjfonts}
\usepackage{float}
\usepackage{ifthen}
\usepackage{natbib}
\def\gs{\mathrel{\raise0.35ex\hbox{$\scriptstyle >$}\kern-0.6em
\lower0.40ex\hbox{{$\scriptstyle \sim$}}}}
\def\ls{\mathrel{\raise0.35ex\hbox{$\scriptstyle <$}\kern-0.6em
\lower0.40ex\hbox{{$\scriptstyle \sim$}}}}

\newfloat{inline}{tb}{}
\newcommand{\inlinefigure}[2]
{\begin{inline}
\noindent\begin{minipage}{0.999\linewidth}
\centerline{\includegraphics[width=0.98\linewidth]{#1}}
\end{minipage}\smallskip
\noindent\begin{minipage}{0.999\linewidth}\footnotesize
{\sc Fig.\ \arabic{figure}.\ ---}\ #2\par
\end{minipage}
\end{inline}
}

\ifthenelse{\equal{\value{figure}}{0}}{\addtocounter{figure}{1}}{}

\makeatother

\shortauthors{{\sc Geach et al}}\shorttitle{\sc Distant, dusty
cluster starburst galaxies}

\begin{document}

\title{The nature of dusty starburst galaxies in a rich cluster at
$z=0.4$:\\the progenitors of lenticulars?}

\author{James\ E.\ Geach\altaffilmark{1}, Ian\ Smail\altaffilmark{1},
Sean\ M.\ Moran\altaffilmark{2}, Tommaso\ Treu\altaffilmark{3} \&
Richard\ S.\ Ellis\altaffilmark{4,5}}

\altaffiltext{1}{Institute for Computational Cosmology, Durham
University, South Road, Durham DH1 3LE, UK. j.e.geach@durham.ac.uk}
\altaffiltext{2}{Johns Hopkins University, Department of Physics and
Astronomy, 3400 N Charles St, Baltimore, MD 21218}
\altaffiltext{3}{Physics Department., University of California, Broida
Hall, MC 9530, Santa Barbara, CA 93106} \altaffiltext{4}{Department of
Astrophysics, University of Oxford, Keble Road, Oxford, OX1 3RH, UK.}
\altaffiltext{5}{California Institute of Technology, 1200 East
California Boulevard, Pasadena, CA 91125}

\begin{abstract} We present the results of a {\it Spitzer} Infrared
  Spectrograph (IRS) survey of 24$\mu$m-selected luminous infrared
  galaxies (LIRGs, $L_{\rm IR} > 10^{11}L_\odot$) in the rich cluster
  Cl\,0024+16 at $z=0.4$.  Optically, these LIRGs resemble
  unremarkable spiral galaxies with e(a)/e(c) spectral classifications
  and [O{\sc ii}]-derived star formation rates (SFRs) of $\lesssim
  2$\,M$_\odot$\,yr$^{-1}$, generally indistinguishable from the
  `quiescent' star forming population in the cluster.  Our IRS spectra
  show that the majority of the 24$\mu$m-detected galaxies exhibit
  polycyclic aromatic hydrocarbon (PAH) emission with implied SFRs
  $\sim$30--60\,M$_\odot$\,yr$^{-1}$, with only one ($<$10\%) of the
  sample displaying unambiguous evidence of an active galactic nucleus
  in the mid-infrared. This confirms the presence of a large
  population of obscured starburst galaxies in distant clusters, which
  comprise the bulk of the star formation occurring in these
  environments at $z\sim 0.5$.  We suggest that, although several
  mechanisms could be at play, these dusty starbursts could be the
  signature of an important evolutionary transition converting
  gas-rich spiral galaxies in distant clusters into the passive,
  bulge-dominated lenticular galaxies that become increasingly
  abundant in the cores of rich clusters in the $\sim$4\,Gyr to the
  present day.
\end{abstract} 

\section{Introduction}

%
%
\input{tab1.tex}

Passive lenticular (S0) galaxies make up a large fraction of the
galaxies in clusters in the local Universe (Oemler 1974), but are
conspicuously absent from equivalent environments just 5\ Gyrs ago at
$z\sim0.5$ (Dressler et al.\ 1997; Smail et al.\ 1997; Couch et al.\
1998; Fasano et al.\ 2000; Treu et al.\ 2003). The reverse is true of
star forming spiral galaxies (Butcher \& Oemler\ 1978, 1984; Couch \&
Sharples 1987), and so it has been proposed that these are the
progenitors of local S0s. Although this proposal is still a
contentious issue, recent work has attempted to piece together a
coherent model for the evolutionary history of star forming galaxies
being assimilated from the field into distant clusters (van Dokkum et
al.\ 1998; Treu et al.\ 2003; Moran et al.\ 2005, 2006; Poggianti et
al.\ 1999, 2004, 2006, 2008). These models generally invoke a
modification of the star-formation histories and morphologies of
spiral galaxies in high-density environments in order to replicate the
spectral properties of the cluster population and the respective
build-up and decline of the S0 and spiral cluster populations since
$z\sim0.5$ (e.g.\ Smail et al.\ 1998; Poggianti et al.\ 1999; Moran et
al.\ 2007).  While there are several viable mechanisms to aid in the
morphological transformation of spirals to S0s: harassment,
ram-pressure stripping, merging, threshing, etc.\ (e.g.\ see Gunn \&
Gott\ 1972; Moore et al.\ 1998; Bekki et al.\ 2001), the luminosities
of the galaxies and in particular their bulges indicates the need for
a period of enhanced stellar mass assembly in the spheroids of spiral
galaxies in clusters to transform them into bulge-dominated S0s by the
present day (Balogh et al.\ 1999; Poggianti et al.\ 1999; Kodama \&
Smail\ 2001; Gerken et al.\ 2004).

The simplest method of enhancing the stellar mass in the bulges of
cluster spirals is to invoke a starburst episode within the central
few kpc of these galaxies.  However, until recently there was little
clear evidence for such starburst activity in these systems (e.g.\
Balogh et al.\ 1997; Poggianti et al.\ 1999).  It has only been with
the availability of sensitive mid-infrared surveys of intermediate
redshift ($z\gs 0.2$--0.5) clusters that potential starburst
populations have been detected (Duc et al.\ 2000, 2004; Fadda et al.\
2000; Metcalfe et al.\ 2003; Biviano et al.\ 2004; Coia et al.\ 2004;
Geach et al.\ 2006; Marcillac et al.\ 2007; Bai et al.\ 2007; Fadda et
al.\ 2008; Oemler et al.\ 2008; Dressler et al.\ 2008).  The
mid-infrared luminosities of the cluster members suggest bolometric
luminosities of $L_{\rm IR} > 10^{11}L_\odot$, characteristic of
Luminous Infrared Galaxies (LIRGs).

In a panoramic 24$\mu$m mapping survey with {\it Spitzer Space
  Telescope (SST)} we detected for the first time a population of
LIRGs distributed to large clustocentric radius in two $z\sim0.5$
clusters (Geach et al.\ 2006).  These galaxies' mid-infrared emission
is far higher than implied from the strength of their optical star
formation rate (SFR) tracers such as H$\alpha$ or [O{\sc ii}].  Such
photometric surveys give no clue to the origin of the intense
dust-obscured activity pin-pointed by the mid-infrared emission, which
could either be star formation or AGN-powered.  If the mid-infrared
emission arises from star-formation, then the star-forming regions
must be highly obscured by the carbonaceous and silicate material
generated by massive stars in the final phases of stellar evolution.

Equally, however, it is also possible that these LIRGs could be
powered by AGN, especially given the evidence for enhanced AGN
activity in intermediate redshift clusters (e.g.\ Martini et al.\
2006; Eastman et al.\ 2007).  Therefore to understand the role of the
luminous infrared population in the evolutionary cycle of cluster
galaxies it is critical to quantify the relative importance of
starburst and AGN emission in the mid-infrared population. If this
population can be conclusively shown to represent starbursting
galaxies, then they could be the long-sort candidates for the
progenitors of local massive S0s, seen at a time when they are
assembling the bulk of their stellar mass. Moreover, the intense star
formation in these distant spirals could explain the apparent young
ages ($\sim5$\,Gyr) of stellar populations in the bulges of S0s in the
cores of local rich clusters (Kuntschner \& Davies\ 1998; Poggianti et
al.\ 2001; Mehlert et al.\ 2003; Arag\'on-Salamanca\ 2007; Bedregal et
al.\ 2008).

In this paper we use low-resolution mid-infrared spectra obtained with
the Infrared Spectrograph on-board the {\it SST} to determine the
origin of the mid-infrared emission in a sample of 24-$\mu$m galaxies
selected from one of the two clusters, Cl\,0024+16 ($z=0.39$), studied
by Geach et al.\ (2006).  We use these data to test what fraction of
these galaxies could be powered by AGN (from the strength of the PAH
features, relative to the warm dust continuum: Rigopoulou et al.\
1999; Genzel et al.\ 1998) -- this is critical for quantifying the
true level of star formation in the 24$\mu$m cluster population. Our
second objective is to determine how the cluster LIRGs'  star formation histories
can be linked with the relatively `quiescent' star forming
population in the clusters. Finally, we examine the potential importance of the
mid-infrared sources as a feeder population for passive lenticular
galaxies found in the cores of local rich clusters. We attempt to
reconcile these observations with the `fossil record' of cluster
galaxy evolution: do they agree with information gleaned from detailed
studies of the star formation histories of S0s in local massive
clusters?

Our study focuses on 24$\mu$m-selected galaxies in Cl\,0024+16.
This is one of the best studied clusters at intermediate
redshifts, and is one of the original `Butcher \& Oemler' clusters
(Butcher \& Oemler 1978, 1984), shown to have a large `blue fraction'
of star forming galaxies (e.g.\ Dressler et al.\ 1997). The cluster
also appears to be in the late stages of a line-of-sight minor merger
that occurred $\sim$3\,Gyr ago (Czoske et al.\ 2002) -- although it is
not clear how this dynamical disturbance has affected the star
formation histories of galaxies in the respective components. The
large observational investment in studies of this cluster,
including deep ground-based optical imaging, high-resolution {\it HST}
imaging studies (Treu et al.\ 2003; Kneib et al.\ 2003); panoramic
H$\alpha$ imaging (Kodama et al.\ 2004); {\it GALEX} ultra-violet
imaging (Moran et al.\ 2006); mid-infrared {\it SST} MIPS and IRS
coverage (Geach et al.\ 2006) as well as wide-field optical
spectroscopy (e.g.\ Czoske et al.\ 2001; Moran et al.\ 2007) results
in an exquisite data set with which to explore the properties of the
star-forming population in this cluster.

Moran et al.\ (2006) investigated the (rest-frame) far-UV (FUV)
properties of so-called `passive spirals' in Cl\,0024+16 (galaxies
with spiral morphologies but with no apparent on-going star
formation). Their red FUV-optical colors (compared to the general
spiral population) suggests that star formation has recently been
truncated in these galaxies. Moran et al.\ (2007) favour a scenario
where gas is prevented from cooling onto the discs of these galaxies,
thus preventing new star formation: `starvation'. It has been argued
that these galaxies are in a transitory phase between active spiral
and passive lenticular galaxy, with gentle truncation of star
formation on time-scales of $\sim1$\,Gyr allowing for residual star
formation during a longer morphological transformation. Yet is still
not clear whether typical spirals' luminosities are sufficient to
build up the brightest lenticulars found in local clusters (Poggianti
et al.\ 1999; Kodama \& Smail 2001).  Thus, part of the motivation of
this work is to understand where the LIRGs fit into this evolutionary
scenario, and whether these luminous galaxies can contribute
significantly to the passive, red color-magnitude sequence in local clusters.
  
In \S2 we describe the IRS observations and data reduction, and
present the key results in \S3. In \S4 we interpret our observations
in terms of the potential evolutionary connection to local S0s. \S5
concludes and summarizes this work.  Throughout we will assume a
cosmological model with $\Omega_{\rm m}=0.3$, $\Omega_\Lambda=0.7$ and
$H_0 = 70$\,km\,s$^{-1}$\,Mpc$^{-1}$.  Unless otherwise stated, we
quote magnitudes on the Vega system.

\section{IRS 7--20$\mu$m spectroscopy}

The 12 galaxies studied with IRS were selected from the 24$\mu$m MIPS
survey of Cl\,0024+16 which is one of the two intermediate clusters
studied by Geach et al.\ (2006).  These 12 galaxies have optical
spectroscopic redshifts confirming them as members of the cluster
Cl\,0024+16 based on the extensive spectroscopic surveys of this
cluster (Czoske et al.\ 2001; Moran et al.\ 2007). In the target
selection, we imposed a 24$\mu$m flux limit of $S_{24} > 0.6$\,mJy
after assuming a spectral shape of $\nu^{-1.5}$ to the underlying
continuum, allowing us to estimate the expected continuum flux density
at 10.9$\mu$m (the PAH\,7.7$\mu$m feature is redshifted to this
wavelength at $z\sim0.4$).

%
%
 \inlinefigure{f1.ps}{Mid-infrared spectra for 24$\mu$m-selected galaxies in
    Cl\,0024+16 (offset in flux density for display purposes). We show
    one example of the fit resulting from the spectral decomposition
    from {\sc pahfit} (spectrum c) clearly showing the good agreement
    with the data ($\chi^2 \sim 1$). We label the main spectral
    features -- most galaxies exhibit strong PAH emission at
    6.2--12.7$\mu$m (note that PAH\,12.7$\mu$m is blended with [Ne{\sc
      ii}]$\lambda$12.813$\mu$m, resulting in the sharp red wing of
    that line). One galaxy is undetected in either SL1 or LL2 modules
    and we do not show that spectrum.  In addition, three galaxies
    have no detectable continuum in the SL1 module and so we only show
    the $>10\mu$m (LL2) portions.  These have weak PAH emission and
    are amongst the faintest of our 24$\mu$m sample. One galaxy is
    classified as an AGN (h), with a red continuum and weak PAH
    emission (a bright X-ray counterpart confirms this
    classification). One of the IRS observations (spectrum a) shows
    two galaxies in close proximity, both revealing two starburst
    spectra in the SL1 module.  We conclude that the majority of
    24$\mu$m-detected galaxies in Cl\,0024+16 are dusty starburst
    galaxies.}\addtocounter{figure}{1}

The 24$\mu$m flux limit, and cluster membership were the only
selections we imposed, and we stress that our sample comprises 12 of
the 13 galaxies with $S_{24} > 0.6$\,mJy and $|\Delta v| <
4500$\,km\,s$^{-1}$ (with respect to the cluster redshift) detected in
our MIPS survey. We note that the sample has a spread in clustocentric
radius representative of the whole MIPS-detected cluster population,
with $r_c \sim 1$--5\,Mpc (see \S3 for more details).

The observations were executed in the Cycle 3 GO project \#\,30621 in
August 2006 and August 2007.  IRS was used in fixed single staring in
modes Short Low 1 (SL1, 7.4--14.5$\mu$m) and Long Low 2 (LL2,
14--21.3$\mu$m), with peak-up acquisition on a nearby 2MASS selected
star (HD\,2292 was a suitable peak-up object for all of our targets).
After peak-up and slew the exposure time is divided into
$5\times60$\,s cycles in SL1 and $20\times120$\,s cycles of LL2,
yielding a total on-object integration time of $\sim5.5$\,ks. The
infrared backgrounds were dominated by Zodiacal Light in each module,
and are in the `medium' category.

We use data products generated by the {\it SST} IRS
pipeline\footnote{{\tt http://ssc.spitzer.caltech.edu/irs/}}. This
processing involves dark subtraction, linearity correction, ramp
correction and flat fielding. Backgrounds were removed by subtracting
two different nod positions, and the 1D spectra optimally extracted
using the latest available release of the {\sc spice} software package
(v2.0.5). This took into account uncertainty and mask images, and the
resulting spectra from each nod position were co-added, rejecting
outliers. Note that all of our targets are treated as point sources
(the width of the IRS slit is 10.5$''$).


%
%
\setcounter{figure}{1}
\begin{figure*} \vspace{3mm}
\centerline{\psfig{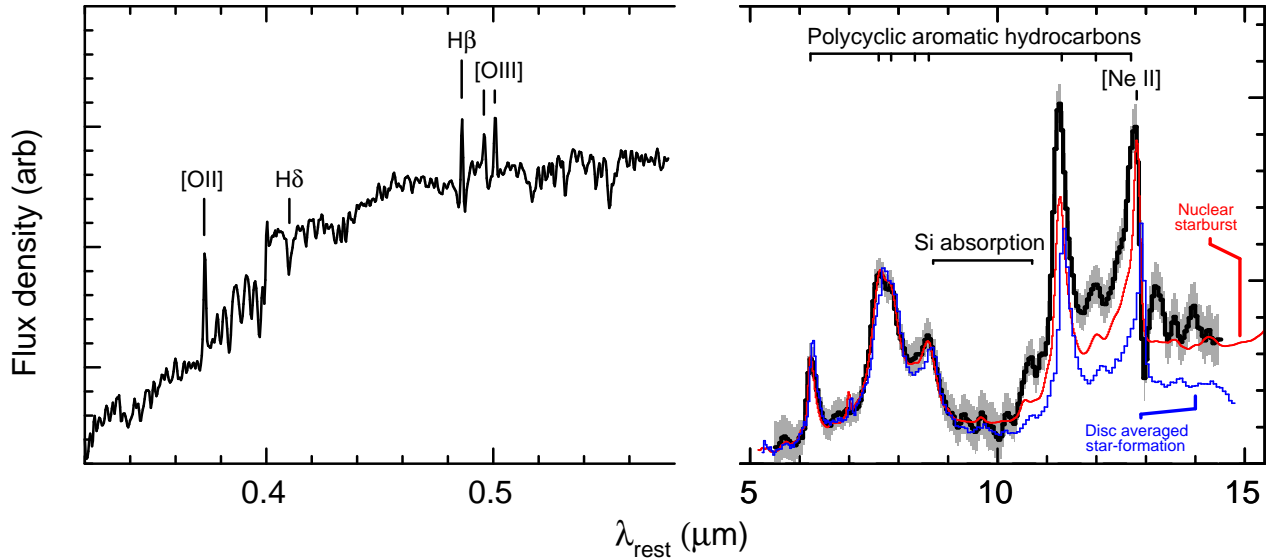}}
\caption{Composite starburst spectrum for cluster LIRGs, shifted to
the rest-frame. The left-hand panel shows the composite optical
spectrum (normalized to unity flux density at 3727\AA). The main
features of the optical spectrum is moderate [O{\sc ii}] emission
[EW([O{\sc ii}])$ = 10.5\pm1.0$\AA] and relatively strong H$\delta$
(and Balmer) absorption [${\rm EW(H\delta)} = 4.6\pm0.9$\AA] :
classified as e(a) in the nomenclature of Dressler et al. (1999), and
other works. The right-hand panel shows the composite (starburst)
mid-infrared spectrum (normalized to unity 7.7$\mu$m, with the shaded
region representing the 1$\sigma$ uncertainty). We compare to the
composite nuclear starburst spectrum of Brandl et al. (2006), and a
composite `disc averaged' spectrum of star forming galaxies from the
SINGS sample (Smith et al. 2007), see \S3. The spectra are all very
similar at $\lambda<10$$\mu$m, but there is much better agreement with
the nuclear starburst spectrum, especially at longer wavelengths. If
the spectral features (silicate absorption, continuum strength, line
ratios, etc.) are connected to the properties of the star forming ISM,
then we suggest that the mode of star formation in the cluster LIRGs is
similar to local nuclear starbursts, as opposed to the relatively
quiescent star formation seen in galactic discs. See \S3 for more details.}
 \end{figure*}

\section{Results}

\subsection{Mid-infrared spectra}

The spectra are presented in Figure~1, and measurements are summarized
in Table~1. Eleven of the twelve targets are well detected in the LL2
module, with the brightest eight also showing detectable continuum in
the shorter wavelength SL1 module.  Only one of the galaxies
(MIPS\,J002624.3) was not detected in either IRS module. We note that
this galaxy is at the faint end of the 24$\mu$m-selected sample and in
addition its optical spectrum shows a passive (k) spectral type, with
no nebular emission, but a strong 4000\AA\ break, with $D_n(4000)=2.3$
(Czoske et al.\ 2001). Thus, its lack of any mid-infrared line
emission may simply reflect that fact that its mid-infrared emission
originates from an AGN with a faint 5--15$\mu$m continuum.  We also
note that other LIRGs with similar 24$\mu$m flux densities
($S_{24}\sim0.6$--0.7\,mJy) are only detected in the LL\,2 module
(Table~1, Figure~1).

\subsection{Star formation versus AGN}

It is immediately clear from Figure~1 that the majority of the
galaxies show PAH emission and this is convincing evidence that the
bulk of the 24$\mu$m selected galaxies in Cl\,0024+16 are dusty
starbursts. Using the line-to-continuum starburst/AGN classification
scheme of Rigopoulou et al.\ (1999) these galaxies would have
$(l/c)\sim2.4$, whereas AGN have $(l/c)<1$ due to the low contrast of
PAH features above the continuum (see also Genzel et al.\ 1998). Half
of the sample shows strong and unambiguous PAH emission at 7.7 and
11.3\,$\mu$m (as well as 6.3, 8.6 and 12.8\,$\mu$m).  In the fainter
sources, e.g.\ MIPS\,J002637.3 or MIPS\,J0002620.2, continuum is only
detectable in the LL2 module, but 11.3$\mu$m PAH is still visible.

Only one galaxy shows unambiguous evidence for AGN-dominated mid-infrared
emission. MIPS\,J002636.3 (Figure~1h) has very weak PAH emission (low
significance 7.7 and 11.3$\mu$m emission), a strongly red continuum
and tentative detection of the [Ne{\sc v}] emission line. The ratio of
line strengths [Ne{\sc v}]/[Ne{\sc ii}] can be used as a diagnostic of
starburst/AGN contribution (Sturm et al.\ 2002), with [Ne{\sc
v}]/[Ne{\sc ii}] generally in the range $\sim0.8$--2 for AGN dominated systems,
compared to strict upper limits for starburst dominated systems of
$<0.01$ (Sturm et al.\ 2002; Verma et al.\ 2003). We do not detect
[Ne{\sc ii}] at 12.8$\mu$m in this source, suggesting
[Ne{\sc v}]/[Ne{\sc ii}]\,$\gs 1.6$, and only very weak PAH
7.7$\mu$m which suggests that this object is almost certainly
dominated by an AGN. Most convincingly, however, inspection of
archival {\it Chandra} X-ray imaging of the cluster reveals that this
galaxy is coincident with a bright X-ray counterpart, with $L_{\rm
0.5-2\,keV} = 2\times 10^{43}$\,erg\,s$^{-1}$ (Soucail et al.\ 2000)
confirming this classification. Interestingly, its optical spectrum
does not show obvious evidence for nuclear activity, but instead shows
post-starburst (k+a) signatures, with no [O{\sc ii}] emission, but
${\rm EW(H\delta)} = 4.8\pm2.3$\AA\ suggesting that an episode of
star-formation has ended within $\sim 1$\,Gyr (Poggianti \& Wu\ 2000)
-- although we caution that the nebular emission lines might be very heavily
extincted. Evidence of weak PAH emission in the mid-infrared spectrum
might hint that there is some level of on-going star-formation.
However, it is likely that the bulk of the mid-infrared emission is
coming from the AGN: the ratio $L_X/L_{IR} \sim 0.03$ measured for
this galaxy is consistent with the infrared-X-ray properties of local
AGN (Alexander et al.\ 2005).

The discovery of just a single AGN-dominated source in our sample of
twelve 24$\mu$m selected cluster members indicates that the vast
majority, $\gs75$\% (including non-detections and ambigious cases), of
the bright 24$\mu$m population in Cl\,0024+16 have mid-infrared
emission dominated by star formation.  This confirms the validity of
the, previously untested, assumption of a dominant starburst component in this
population used in the analysis of Geach et al.\ (2006) and all other
papers analysing mid-infrared surveys of distant clusters to date.  For the remainder
of this paper we concentrate on understanding the properties of the
dusty starbursts in our sample and their significance for models of
cluster galaxy evolution and the total star formation rate in the
cluster.

\subsection{Composite spectrum and comparison to optical properties}

Since we have precise redshifts for these starburst-classified
galaxies (from their optical spectra), we can co-add the rest-frame
IRS spectra into a composite `star forming' spectrum shown in
Figure~2. This yields the mid-infrared spectrum of a typical galaxy
with better signal to noise than in the individual spectra, and
clearly shows strong PAH emission at 6.3, 7.7, 8.6, 11.3 with a
PAH/[Ne{\sc ii}] blended line at 12.7$\mu$m.  For comparison, we show
the equivalent composite optical spectrum for these galaxies (Moran et
al.\ 2007; Czoske et al.\ 2001). It resembles a typical spiral galaxy,
with signatures of dust obscured star formation -- it would be
classified as e(a) in the nomenclature of the {\it Morphs}
collaboration (Dressler et al.\ 1999).  It has relatively weak [O{\sc
  ii}] emission, implying on-going un-extinguished star formation with
rates of SFR([O{\sc ii}])$\lesssim$2\,$M_\odot$\,yr$^{-1}$. The
moderate H$\delta$ absorption suggests that there has been massive
star formation on a time scale of $\lesssim$1\,Gyr (Poggianti et al.\
1999). Turning to the full sample of 24$\mu$m-detected members of
Cl\,0024+16, of all the $S_{\rm 24\mu m}>0.2$\,mJy galaxies in
Cl\,0024+16 with spectral classifications from Czoske et al.\ (2001)
and Moran et al.\ (2007), 53$\pm$12\% are e(c), 29$\pm$9\% are e(a),
$5\pm4$\% are k+a (including the AGN described above) with the
remainder classified as k (passive).  Hence, the majority of them fall
into the moderately star-forming classes, e(c) -- characteristic of
normal spirals or e(a) -- a classification that has been argued as an
optical signature of a dusty starburst (Poggianti \& Wu\ 2000).  We
note that our IRS sub-sample (i.e.\ those bright in the MIPS waveband,
$S_{24}>0.6$\,mJy), has a slightly higher fraction of e(a)
galaxies: $40\pm20$\%. This might hint that the optically identified
dusty starbursts (i.e.\ the e(a) class) may be biased towards the most
active galaxies (see also Dressler et al.\ 2008), and lower (but still
significant) levels of star formation are hidden in the more numerous,
but less extreme, e(c) classified MIPS cluster members.

\addtocounter{figure}{1}
\inlinefigure{f3.ps}{Calibration of total infrared (TIR, 8--1000$\mu$m)
luminosities from the luminosity of the 7.7$\mu$m blended PAH feature
when measured with {\sc pahfit} (modeling PAH emission with a set of
Drude profiles, and continuum as a superposition of modified
black-bodies). We use the catalog of H{\sc ii} galaxies from the
SINGS survey (Smith et al.\ 2007) to demonstrate that there is a
strong correlation between the PAH line strength and the total
bolometric output that spans over several orders of magnitude in
activity. We also show the positions of our LIRG sample, with $L_{IR}$
derived from our previous extrapolation from 24$\mu$m luminosities
(Geach et al.\ 2006).  We fit a linear trend to a the $L_{IR} >
10^{8}$\,$L_\odot$ data, which agrees very well with both the
low-luminosity, local SINGS galaxies, and our more active LIRGs at
$z=0.4$ (the SINGS sample has been corrected for aperture losses). As
a guide, we show the range of fits when the luminosity distribution is
bootstrap re-sampled, and we take this as a conservative estimate of
the typical uncertainty in our calibration. Note that the right-hand
axis shows the conversion from infrared luminosity to SFR assuming the
relationship in Kennicutt\ (1998).  }

The mid-infrared spectra of these galaxies (Figures~1 \& 2) reveals
that these galaxies are actually undergoing significantly more star
formation than the optical nebular lines would suggest. We compare the
cluster LIRGs' composite mid-infrared spectrum to a composite spectrum
extracted from the nuclear regions of a sample of local starbursts
(Brandl et al.\ 2006) and a `disc averaged' composite spectrum
constructed from a selection of SINGS galaxies (Smith et al.\
2007)\footnote{Selected such that their angular size is most
  comparable to the IRS slit, and therefore approximate to
  galaxy-integrated spectra.}. Although there is broad agreement
between the features in all the spectra (especially at $\lambda <
10$\,$\mu$m), our cluster LIRGs show stronger emission at longer
wavelengths, $\lambda> 10\mu$m, than either of the two local templates
-- although they are more similar to the the local nuclear starburst
spectrum.  We have attempted to improve the match between the local
templates and the cluster LIRGs by modifying the SINGS (or local
nuclear starburst) composite spectrum through the addition of a
power-law continuum (mimicking a contribution from hot dust), combined
with Galactic centre extinction (Chiar \& Tielens\ 2005).  Although
such a model allows us to better match the continuum in the cluster
LIRGs' spectra at $>10\mu$m, it still fails to match the PAH line
ratios (the $11.3/7.7\mu$m PAH line ratio is particularly sensitive to
the gradient in the reddening curve, $A_{11.3}/A_{7.7}\sim2$ in this
model).  To achieve a good match, we would have to enhance the
$11.3\mu$m and $12.7\mu$m PAH features in the SINGS spectrum, however
the physical motivation for such a change is uncertain and complex
(related to metallicity, grain ionisation fraction, dust size
distribution, etc., see Draine \& Li\ 2001; Calzetti et al.\ 2007) and
so we feel it is unjustified here.

Nevertheless, we take the
similarity between the nuclear starburst spectrum and our
galaxy-integrated spectrum as suggestive  that the cluster
LIRGs are dominated by a similar (nuclear) mode of star formation.
The best local analogue of the cluster LIRGs from the Brandl sample is
NGC\,3310. This UV- {\it and} IR-bright starburst, has morphological
signs that it has recently undergone a minor merger, and (like many
starbursts) has star-formation concentrated in a circumnuclear ring
(Conselice et al.\ 2000). The mixture of bright UV and infrared
emission suggests that the star-forming regions of this galaxy are
differentially obscured according to age (e.g.\ Poggianti et al.\ 2001)
-- this might explain the weak spectroscopic signature of a
significant population of O stars (Leitherer et al.\ 2002).  In the
next section we use the PAH emission features to estimate total
infrared luminosities, and thus more reliable SFRs for cluster LIRGs.
 
\subsection{Star formation rates}

In the presence of dust, infrared tracers of star formation are more
reliable than optical calibrations. In our previous photometric study
(Geach et al.\ 2006) we estimated the total 8--1000$\mu$m emission
from cluster LIRGs by assuming a simple ratio between the observed
24$\mu$m continuum emission and total infrared (TIR) luminosity based
on a library of dusty starburst SEDs characterized by a power-law
distribution of dust mass (Dale \& Helou [D\&H]\ 2002). This result
was consistent with the local relation between 15$\mu$m and
far-infrared luminosity (Chary \& Elbaz\ 2001), but using our
new IRS spectra we can now improve the choice of the appropriate
model SED and hence the predicted bolometric luminosities.  The
{\it Spitzer} 24$\mu$m MIPS band traces the integrated emission at
$\lambda_{0}\sim 15$--19\,$\mu$m in the cluster member's
rest-frame. This region of the mid-infrared spectra of galaxies
contains both PAH emission, continuum emission and silicate absorption
(not modeled in the D\&H SEDs), and so the spectral variation inherent
to star-forming galaxies (e.g.\ Draine et al.\ 2007; Smith et al.\
2007) could result in significant over or under-estimation of the
total SFR, depending upon the 
precise mix of absorption and emission in the MIPS band. 
Alternatively, if the galaxy contains an AGN, we could
erroneously overestimate the SFR -- thus one of the main objectives of
this work was to obtain an unambiguous, accurate SFR estimate for
cluster LIRGs.

The new IRS spectra provide a more accurate picture of the nature of
the mid-infrared emission, which is superior for deriving SFRs. This is not
only because the emission features are more closely correlated with
the intensity of the ionizing UV photons from massive stars, but also
since the total mid-infrared emission can be dominated by these
features (Smith et al.\ 2007). Our estimates of the PAH line
luminosities are derived by fitting the 5--15$\mu$m rest-frame spectra
with {\sc pahfit}\footnote{{\tt
http://turtle.as.arizona.edu/jdsmith/pahfit.php}}, the spectral
decomposition code of Smith et al.\ (2007). This method models the PAH
dust features as Drude profiles (appropriate for harmonic oscillators
-- the C--H bonds in PAH molecules). The derived PAH\,7.7$\mu$m and
PAH\,11.3$\mu$m line fluxes are presented in Table~1.

Although the PAH features are thought to arise directly from the
excitation of molecules by UV emission associated with massive star
formation, the complex physics of the macromolecules means that
directly linking PAH line luminosity to star formation rates remains a
challenge (e.g.\ Wu et al.\ 2005). An alternative method is to use the
PAH emission as a proxy for the total infrared emission; there is a
tight correlation between PAH line and total infrared emission in
galaxies spanning a wide range of redshift and luminosity (Peeters et
al.\ 2004; Brandl et al.\ 2006; Schweitzer et al.\ 2006; Pope et
al.\ 2008). Here we adopt the PAH 7.7$\mu$m line to estimate the
bolometric luminosity\footnote{We refer to the `PAH\,7.7$\mu$m'
feature -- in fact, this is a blend of several PAH emission lines, and
the luminosities reported here are the total for the blended feature,
see Smith et al.\ (2007) for more details on the spectral
decomposition. The reader should also that {\sc pahfit} overestimates line fluxes by a
constant factor $(1+z)$, but our spectra were fit in their
rest-frame.}.  To calibrate $L_{\rm PAH\,7.7}$ to total infrared
(8--1000$\mu$m) luminosities, we compare $L_{\rm PAH7.7}/L_{\rm
8-1000}$ for H{\sc ii}-dominated galaxies (i.e.\ star-forming) selected
from the SINGS catalog (Smith et al.\ 2007). These galaxies span
several orders of magnitude in both total infrared and PAH luminosity
with a clear linear trend between the two (Figure~3). We find:
\begin{equation} \log~(L_{IR}/L_\odot) = (1.01\pm 0.01)\log~(L_{\rm
PAH 7.7}/L_\odot) - (1.27\pm 0.05)
\end{equation} Note that this differs in part from calibrations presented in
other works (e.g.\ Pope et al.\ 2008), since here we are using the
Drude profile to estimate line luminosities, whereas other studies
have used alternative methods to estimate the PAH line strengths
(spline fitting, Gaussian profiles, etc.) which tend to result in
systematically lower luminosities.  To derive a SFR from $L_{IR}$, we
adopt the calibration of Kennicutt (1998): ${\rm SFR(IR)} =
4.5\times10^{-44}L_{IR}/{\rm (erg~s^{-1})}$. The resulting SFRs are
listed in Table~1.  To compare to our previous estimate of the
24$\mu$m-derived SFR in these galaxies, we apply the same 24$\mu$m/TIR
calibration as in our previous work (Geach et al.\ 2006) and find 
reasonable
consistency with the PAH\,7.7/TIR calibration (Figure~3), with $\left<
{\rm SFR(PAH\,7.7\mu m)}/{\rm SFR(24\mu m)}\right> = 0.77\pm0.29$.

Obtaining IRS spectra of a sample of LIRGs in Cl\,0024+16 also
provides us with a method to more accurately determine the SFRs of all
24$\mu$m-detected ($S_{\rm 24} >0.2$\,mJy) galaxies in this cluster,
since we can determine the best match local analogue SED to derive
$L_{IR}$, instead of relying on a suite of SED types (although it is
still possible that there is some variation amongst the LIRGs). While
we lack spectral coverage beyond $\lambda_0 \sim 15\mu$m, the
24$\mu$m photometry provides a good estimate of the continuum level at
$\lambda_0\sim17\mu$m, thus improving any comparison where hot dust
may be important. The best matching SED in the D\&H library
corresponds to $\alpha=1.6875$\footnote{The D\&H library of SEDs are
characterized by $\alpha$, the exponent in the power-law distribution
of the mass of dust over the intensity of the dust heating radiation
field: ${\rm d}M_{\rm d}(U) \propto U^{-\alpha}{\rm d}U$. See Dale \&
Helou\ (2001) for more details.}, with an {\it IRAS} far-infrared
color of $f_{60}/f_{100} \sim 0.75$.

When this SED is used in the transformation from 24$\mu$m to
total infrared emission for all the 24$\mu$m-detected members of
Cl\,0024+16, and including a conservative 25\% correction for AGN
contamination in the total 24$\mu$m luminosity, we estimate a
total SFR of $735\pm150$\,$M_\odot$\,yr$^{-1}$ (limiting to within
2\,Mpc of the core; the uncertainty is derived by
bootstrap re-sampling the LIRG flux distribution). This compares to
the original estimate of $\sim 1000\pm 210$\,$M_\odot$\,yr$^{-1}$ by
Geach et al.\ (2006) and the estimate of $\sim
220$\,$M_\odot$\,yr$^{-1}$ derived over the same luminosity range from
an H$\alpha$ narrow-band imaging survey by
Kodama et al.\ (2004).  The change in our estimate of the total SFR in
the cluster is driven by our new (albeit conservative) estimate of the
likely AGN contamination, measured with the present IRS observations.
Despite the small AGN correction, our IRS observations confirm that
the dust-obscured mode of star formation dominates the most intense star
forming events in the cluster population and likely produces the bulk
of the stars formed in cluster galaxies at $z\sim 0.5$.

\subsection{Environment}

In Figure\ 4 we present the velocity and radial distribution of
24$\mu$m-selected ($S_{\rm 24} > 0.2$\,mJy) galaxies in Cl\,0024+16
compared to the general cluster population. As mentioned in \S1,
Cl\,0024+16 is a line-of-sight merger, visible clearly in Fig~4 by the
distinct foreground component of galaxies, blue-shifted by
$\sim3000$\,km\,s$^{-1}$ with respect to the main cluster at $z=0.395$
(Czoske et al.\ 2002). Note that the large fraction (similar to the
main cluster) of 24$\mu$m-detected galaxies in the foreground
component is interesting: if these galaxies were members of this
cluster prior to the collision, then they must have been subjected to
a high-speed pass through the ICM of the main cluster. As evidenced by
their high SFRs, they did not lose their internal gas reservoirs
during this episode -- implying that they have not been
ram-pressure stripped. Alternatively, these starburst galaxies may have been
accreted onto the foreground component some time after the collision.

%
%

\addtocounter{figure}{1}
 \inlinefigure{f4.ps}{Velocity-radial distribution of
    24-$\mu$m selected ($S_{\rm 24} >0.2$\,mJy) galaxies in
    Cl\,0024+16 compared to the general cluster population (MIPS
    detections are circled. Note that our previous MIPS survey
    excluded the central $\sim2.5'\times2.5'$ due to GTO constraints
    -- our analysis thus concentrates on the region outside of the
    core). The velocity distribution shows that the cluster is
    actually composed of two components: the main cluster at
    $z=0.395$, and a smaller `foreground' component at $\Delta
    v\sim-3000$\,km\,s$^{-1}$. It is thought that this is the remnant
    of a line-of-sight group-cluster collision that occurred some
    $\sim3$\,Gyr ago (Czoske et al.\ 2002). The curved lines show the
    escape velocity for the mass enclosed within a given radius
    (assuming $M=8.7\times10^{14}M_\odot$ [Kneib et al.\ 2003] and an
    isothermal potential). Mid-infrared luminous galaxies are
    distributed out to large clustocentric radii with a velocity
    dispersion similar to the general spiral population -- there is no
    convincing evidence to suggest that the luminous obscured
    starbursts inhabit significantly different local environments than
    the majority of the star forming population in Cl\,0024+16,
    although several LIRGs (like the lower luminosity spirals) belong
    to small groups.}

 \begin{figure*}
\centerline{\psfig{file=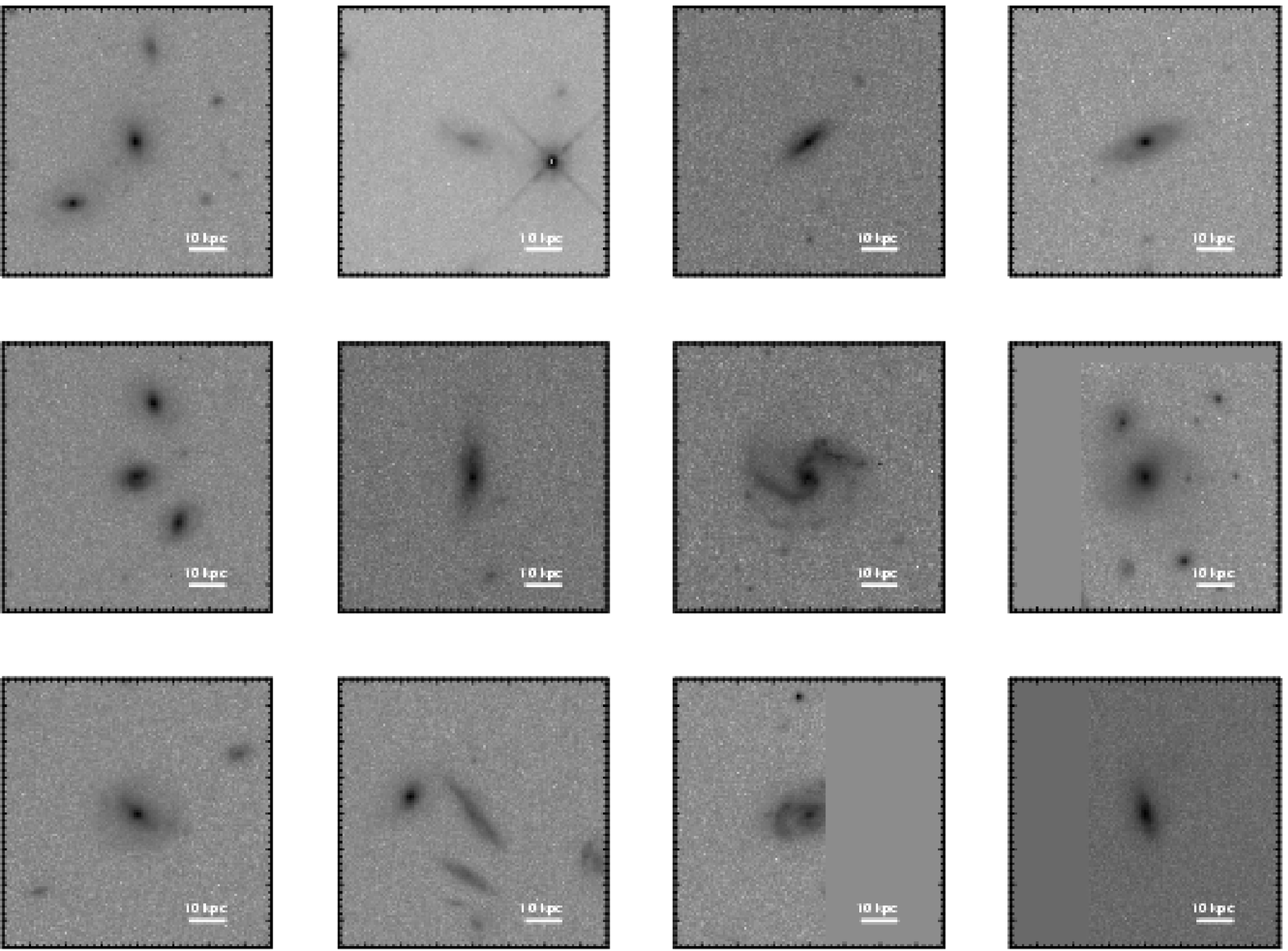,width=0.95\textwidth}}
 \caption{{\it Hubble Space Telescope} WFPC2 (F814W, rest-frame {\it
V}-band) images of 12 $S_{\rm 24} > 0.2$\,mJy members of
Cl\,0024+16. We see a range of morphologies, although many galaxies appear to be relatively early-type spirals. All images are $15''\times15''$ ($\sim 80\times80$\,kpc),
and ordered from left to right, top to bottom in increasing infrared
luminosity. The images have been smoothed with a Gaussian filter
(kernel FWHM 250\,pc) and are logarithmically scaled.  }
\end{figure*}

Focusing our attention on the main cluster, the 24$\mu$m-detected
members have similar velocity and radial distribution to non-detected
spirals, with an average clustocentric radii of $r_c =
2.7\pm1.0$\,Mpc, $\sigma = 857\pm426$\,km\,s$^{-1}$ and $r_c =
2.4\pm0.9$\,Mpc, $\sigma = 969\pm214$\,km\,s$^{-1}$ measured for the
two populations respectively\footnote{We only consider cluster members
  in the main cluster component and with radii $>$0.85\,Mpc, matching
  the region covered by our previous MIPS survey. Velocity dispersions
  are calculated from the bi-weight scale of the distribution (Beers,
  Flynn \& Gebhardt 1990) with an uncertainty derived from bootstrap
  re-sampling the velocity distribution.}.  For comparison, the E/S0
galaxies have a slightly lower velocity dispersion, $\sigma =
776\pm77$\,km\,s$^{-1}$ (but a similar radial distribution if one
excludes the central region not covered by the Geach et al.\ (2006)
MIPS sample, $r_c = 2.4\pm1.0$\,Mpc). This kinematic difference can be
attributed to the fact that the MIPS-detected population and the less
active spiral galaxies predominantly stem from accretion (infall) from
the field and are not yet fully virialised.

We find no evidence that LIRGs have significantly different local
environments to the general cluster population, but (like the general
spiral population) several LIRGs appear to be members of small
groups. One of these groups contains two 24$\mu$m detected members,
and is at a large relative velocity compared to the main component
($\sim2500$\,km\,s$^{-1}$). Incidentally, one of these galaxies is the
AGN in our IRS sample (MIPS\,J002636.3, Table\,1h, Fig\,1h). This is
probably part of a small bound group of galaxies falling into the
cluster. Several more 24$\mu$m-detected members are part of a large
sub-structure to the north-west of the main cluster core. It is
possible that some of the triggering of starburst (and AGN) activity in
Cl\,0024+16 occurs through interactions within these
small bound groups. Evidence of this behaviour can be seen
in Fig~1a: two cluster members were detected in the IRS SL1 mode (they
were at a close separation and both fell on the slit): both are
bright enough in the mid-infrared that we can see that they have very
similar, dusty starburst spectra.

If we parameterise the galaxies' local environment with the
Dressler--Shectman (DS) statistic (Dressler \& Shectman 1988), we find
no statistical difference between the starbursting LIRGs and
`quiescent' spirals -- which display similar local environments. However, as noted
by Moran et al.\ (2007) and described above, the overall level of
sub-structure in Cl\,0024+16 could have significant implications for
the evolution of the star-forming population, with galaxies much more
likely to have been subject to `pre-processing' in small bound groups
prior to cluster assimilation.  We interpret the similarity in environmental
properties between 24$\mu$m-selected and normal spirals in Cl\,0024+16
as
an indication that the LIRGs are simply the bright tail of the star forming
population in this cluster, whose enhanced activity does not reflect a unique environment, but rather points to these galaxies being particularly gas-rich.
 
\subsection{Morphologies}
 
An important aspect of the proposed evolutionary link between cluster
spirals and S0s is the necessity for a morphological
transformation. Luckily, a fraction of the LIRG members of Cl\,0024+16
fall within the sparse {\it Hubble Space Telescope} WFPC2 mosaic of
the cluster (Treu et al.\ 2003), and therefore can be classified
morphologically.  From our IRS sample, we have WFPC2 imaging of three
galaxies and we supplement this small sample in Figure~5 with the
remainder of 24$\mu$m-selected cluster members with $S_{24} >
0.2$\,mJy that are covered by the {\it HST} survey to demonstrate the
range of morphologies for the population.  We find that 80\% of the
LIRGs with classifications are designated as early-type spirals (Sab
or earlier), although several galaxies exhibit obvious spiral arm
structure and most have prominent disk components. Of the full sample,
8 galaxies have bright bulges and/or nuclear point sources, and 5 have
evidence for recent or on-going interactions (we cannot rule out
advanced mergers in some cases); while several galaxies exhibit
slightly disturbed morphologies. This might be evidence of recent
processing in the cluster environment or in small groups (e.g.\
harassment, high-speed encounters in the cluster potential, etc.).  As
described above, LIRGs appear not to have significantly different
local environments to the general spiral population, however evidence
of disturbed morphologies hints that local dynamical disturbances have
a role in triggering star formation over all luminosities (and in the
case of LIRGs potentially help in completing the spiral-to-S0
morphological transformation by subsequently erasing any remaining
spiral structure).

\section{Discussion}

The results presented above highlight the fact that our understanding
of the evolutionary history of cluster galaxies is still
incomplete. The vigorous star formation which our IRS spectra have uncovered
in the cluster LIRG population means that these galaxies are
potentially good candidates to evolve into the lenticular galaxies
that become more abundant in the cores of rich clusters at
$z\lesssim0.5$. In the following discussion, we examine this scenario,
concentrating mainly on the star formation properties and histories of
cluster galaxies in Cl\,0024+16.
 
\subsection{The star-formation histories of cluster LIRGs}

It is clear from studies of generally passive 
galaxy populations in local clusters that the
eventual fate of gas-rich spiral galaxies entering clusters at high
redshift must be the cessation of their star formation. It is not clear
precisely what mechanisms control this evolution: there are a wide
range of processes occurring in cluster environments that 
could potentially affect
the star formation histories of galaxies traversing the cluster
potential (e.g.\ Gunn \& Gott\ 1972, Cowie \& Songaila 1977; Dressler
\& Gunn\ 1983; Bekki 1998; Moore, Lake \& Katz 1998; Moore et al.\
1999).

As we showed in \S3.5, the mid-infrared starburst galaxies share
similar environmental conditions to the less active spirals. It is
therefore reasonable to assume that both populations are subject to
similar environmental processing. This suggests that the enhanced SFRs
seen in the LIRGs owes more to the galaxies' individual properties
than their environment. Although perhaps not true of all clusters, in
terms of the survival of star formation, Cl\,0024+16 appears to exert
a weak influence on infalling spiral galaxies (Moran et al.\ 2007).

In fact, ram pressure stripping can largely be ruled out as an
effective truncation mechanism for the starburst population in
Cl\,0024+16: stripping is only effective within $\sim1$\,Mpc of the
core, and the majority of the mid-infrared sources are outside this
region (Figure~4).  The typical time-scale for galaxies to reach the
core region is $\sim1$--2\,Gyrs, although the actual stripping
time-scale is comparatively short (a few tens of Myr) once the
galaxies reach the highest density regions (see the discussion in Treu
et al.\ 2003 for more details).  However, the lifetime of the
starburst events in these galaxies is likely to be of the order of
100\,Myrs (see \S4.3), thus, by the time LIRGs reach the cluster core,
their bursts will be long-since over, and so (for Cl\,0024+16 at
least) ram-pressure stripping is unlikely to play a dominant role in
modification of LIRGs' star formation histories.  A caveat, as noted
by Moran et al.\ (2007), is that the presence of significant
sub-structure in the cluster could give rise to shocks in the ICM that
might exert significant ram-pressure at comparatively large
clustocentric radius -- potentially responsible for the triggering of
star formation in in-falling groups. If we assume that it is the core
regions where truncation of star formation takes place, how will this
affect the LIRGs?  Most likely, once the LIRGs sink to the bottom of
the potential well, any residual star formation they do have will be
finally extinguished, but their major epoch of star formation will
have already occurred.

Moran et al.\ (2007) propose that spiral galaxies in Cl\,0024+16
experience a gentle `starvation' of their gas reservoirs as they
encroach on the potential well (i.e.\ residual star formation exhausts
gas reservoirs, with no replenishment from cooling onto the disc).
This model has been used to support the scenario where spirals are
gradually transforming into passive lenticulars (or dwarf
spheroidals), but it is not clear that typical starved spirals can
account for the formation of the most luminous lenticulars found in
local clusters. Starvation is evidently irrelevant for LIRGs, since
their gas will probably be converted to stars in a short period
(few 100\,Myrs) compared to the 1\,Gyr truncation timescale of normal
spirals. A natural explanation for the existence of LIRGs in
Cl\,0024+16 is that we are observing starbursts that originated as the
most gas-rich spirals in the field: these galaxies can achieve 
high bulge luminosities at the present-day
without relying on replenishment of their gas
reservoirs from further cooling.

If the starbursts in Cl\,0024+16 simply exhaust their gas reservoirs
unmolested, then it is straightforward to estimate their luminosity
evolution and thus compare to the properties of local cluster
populations. The typical rest-frame $H$-band luminosity of the
Cl\,0024+16 LIRGs is $M_H=-24.5\pm 0.6$\,mag. Assuming complete
cessation of star formation after the burst, and passive evolution
until $z=0$, we estimate a total ($H$-band) fading of $\Delta M_H
\sim2$--2.4\,mag (the range reflects the difference between star
formation histories where (a) all the stellar mass is formed in the
burst, and (b) the luminosity includes a pre-existing older stellar
population formed continuously since $z=5$\footnote{We base these
  estimates on the application of the simple stellar population models
  of Bruzual \& Charlot (2003), assuming a constant Solar metallicity,
  and Chabrier initial mass function.}).  Thus, the LIRGs descendants
will have typical luminosities of $-21.5 \lesssim M_H
\lesssim-23.1$\,mag. This luminosity evolution appears to be
sufficient to place their descendants at the bright-end of the S0
luminosity function in local clusters; for example the median $H$-band
luminosity of morphologically classified S0s in Coma is $M_H =
-22.7$\,mag (J.\,R.\ Lucey, private communication). This is a
conservative estimate, because it is feasible that some residual
`quiescent mode' star formation could endure in these galaxies over
the $\sim4$\,Gyr to the present day.  In contrast, the `starved'
descendants of typical spirals in Cl\,0024+16 are not luminous enough
to populate the bright end of the local S0 luminosity function, and
generally have a final $M_H > -22.7$\,mag (see Smail et al.\ 1998). We
therefore suggest that the `dusty starburst phase' identified by our
joint MIPS/IRS study is a key evolutionary step required to boost the
luminosity of infall spirals if their progeny are to become the most
massive S0s found in local clusters (Poggianti et al.\ 2001; Kodama \&
Smail\ 2001).

\subsection{Bulge formation: circumnuclear starburst?}

While there may be sufficient star formation in cluster LIRGs to match
the luminosities of local S0s, where is this activity occurring within
the galaxies? A key requirement in the conversion of a spiral to a
lenticular galaxy is the enhancement of stellar luminosity in the
galactic bulge compared to the disc.  A critical signature of bulge
growth would be the observation of a circumnuclear starburst, where
the stellar mass in the bulge can be rapidly enhanced without the need
for dramatic migration of the stellar populations. Unfortunately, our
current mid-infrared observations do not have the spatial resolution
to confirm this mode of star formation -- nevertheless, we argue that
a circumnuclear starburst is a natural scenario for these galaxies.

Our evidence is largely circumstantial. As discussed in \S3, the
cluster LIRGs' galaxy integrated spectra closely resemble that of
spectra extracted from the nuclear regions of local starbursts
compared to the disc-integrated spectra of typical local star forming
galaxies (see Figure~2). Given that mid-infrared spectral properties
can largely be driven by the conditions of the ISM environment, we
take this similarity as a compelling hint that the LIRGs mode of star
formation occurs in a circumnuclear environment. Furthermore, the
overall rate of star formation (as measured from the PAH spectral
features) is hard to reconcile with an extended disc. In the local
Universe starbursts tend to be centrally concentrated, given the
elevated gas densities found in the nuclear regions compared to the
extended disc (Kennicutt 1998a). If star formation were distributed
over the discs of these galaxies (and assuming the scaling between
star formation rate surface density and gas surface density holds),
then the average gas surface densities required would be nearly an
order of magnitude larger than that seen in typical star forming discs
(Kennicutt 1998b). In contrast, the gas densities found in typical
compact starburst discs is much more amenable for the observed levels
of star formation.

One other piece of evidence that suggests that the LIRGs have enhanced
star formation in their bulges, compared to their discs comes from
their morphological properties (\S3.6): $\sim80$\% of
24$\mu$m-detected cluster members that are covered by the sparse {\it
HST} WFPC2 mosaic are classified as early type spirals (Sab) or
earlier type (Figure~5). This could imply two things: (a) a large amount
of the morphological transformation from spiral to lenticular is
already complete, and all that remains is an overall boosting of
luminosity, and/or (b) the bulge luminosities of these galaxies is
overwhelming the discs because the current starburst is occurring in
the bulge region (e.g.\ Bendo et al.\ 2002a,b).

Finally, the cluster environment itself could support triggering of
nuclear star formation: at large clustocentric radii, although ram
pressure {\it stripping} is largely ineffective, the
IGM/ISM interaction can still influence star-formation histories
(Tonnesen et al.\ 2007). For example, compression of molecular clouds
in the discs of infalling spirals could serve to funnel material
toward the core (Kronberger et al.\ 2008, also see Byrd \& Valtonen
1990 \& Natarajan et al.\ 2002, 2008). We note however that this
massaging effect could also could trigger star-formation throughout
the galaxy disc (Dressler \& Gunn 1983; Bekki \& Couch 2003). Closer
to the cluster core, where stripping actually takes place, one would
also expect to observe circumnuclear starbursts -- this is the final
(or at least hardest) region of the galaxy to be stripped of gas, and
so any residual star formation is likely to occur there (but see
Quillis et al.\ 2000).

\subsection{Comparison to the fossil record}

One experiment that would lend support for the idea that distant
starbursts are feeding the population of present day lenticulars via
enhanced bulge growth would be signatures of $\sim4$\,Gyr old stellar
populations in the bulges of {\it local} cluster S0s. Spectroscopic
studies of the core regions of early type galaxies in local rich
clusters have shown that a large fraction of S0s have relatively young
stellar populations consistent with star formation within the past
5\,Gyrs, with less luminous S0s showing younger bulge ages (Poggianti
et al.\ 2001; Mehlert et al.\ 2003; Bedregal et al.\ 2008; Kuntschner
\& Davies 1998).  The observed luminosity offset of local S0s from the
spiral Tully-Fisher relationship is also consistent with the fading
scenario described in \S4.1 (e.g.\ Arag\'on-Salamanca\ 2007; M.\ A.\
Norris private communication).

In order to explain the apparent luminosity-dependent distribution of
ages measured for the bulges of local S0s, either the bright end of
the S0 luminosity function is made mainly up from galaxies that formed
in monolithic collapse or wet mergers at high redshift $z\gtrsim0.5$,
or the most intense star-formation episodes responsible for the
formation of massive S0s today are increasingly common at
higher-redshift.  This might be a reflection of the apparent strong
evolution in the number density of (field) LIRGs since $z\sim1$ (Cowie
et al.\ 2004) compared to that of `normal' star forming galaxies
(Lilly et al.\ 1995).  Indeed, we already have hints that this
behavior is seen in cluster environments when one considers the {\it
total} star formation rates of massive clusters of galaxies since
$z\sim1$ (e.g.\ Geach et al.\ 2006). Of course, we should also
consider alternative mechanisms giving rise to the local S0
populations. For example, dry-merging between evolved stellar systems
could also be responsible for the formation of the massive end of the
S0 population (e.g. van Dokkum 2005).  This leads us to the final
point to address: are there sufficient numbers of these distant
galaxies to explain the local abundance of passive lenticular galaxies
in local clusters?

Assuming a simple evolutionary model where the LIRGs transform into
S0s, one can estimate the required abundance of these galaxies in
order to match the local populations (e.g. Smith et al.\ 2005). We
apply the same model as Moran et al.\ (2006) for passive (starved)
spirals in Cl\,0024+16:
\begin{equation} \left(\frac{N_{\rm S0}}{N_{\rm E}}\right)_{z=0} =
\left(\frac{N_{\rm S0}}{N_{\rm E}}\right)_{z=0.4} + \frac{\Delta
t}{\tau} \left(\frac{N_{\rm LIRG}}{N_{\rm E}}\right)_{z=0.4}
\end{equation} Here $\Delta t = 4$\,Gyr (the intervening time since
$z=0.4$) and $\tau = 100$--200\,Myr, a conservative range of expected
lifetimes for the starbursts. We take the Cl\,0024+16 S0:E abundance
from Dressler et al.\ (1997), and apply this to the most recent
spectroscopic catalogue of cluster members (Moran et al.\ 2007) to
estimate $N_E$ in the cluster. Limiting our estimates to the fraction
of $S_{24} > 0.6$\,mJy (or $\log (L_{IR} / L_\odot) \gtrsim 11.2$)
LIRGs bound to the main cluster ($|\Delta v| < 2000$\,km\,s$^{-1}$,
see Fig~4), we find $N_{\rm LIRG}/N_{\rm E} \sim 0.04$, resulting in a
local S0:E ratio of 1.5--2.3 (the range is connected to the span of
LIRG lifetimes in the model). This is consistent with the observed
local S0:E fraction in massive clusters: $N_{\rm S0}/N_{\rm E} =
1.7\pm0.6$ (Dressler et al.\ 1997) and implies that the observed
abundances of LIRGs in $z\sim0.5$ clusters can account for the
demographics of S0s in local massive clusters.

\section{Conclusions and final remarks}

Our previous work uncovered a large population of luminous
mid-infrared sources in the rich cluster Cl\,0024+16. This work
investigated a sub-sample those galaxies in greater detail, using new
{\it Spitzer} IRS spectroscopy specifically their mid-infrared and
optical properties. Our main results can be summarized as follows:
\begin{enumerate}
\item{The LIRGs' mid-infrared spectra confirm that the majority of the
    24$\mu$m-selected galaxies in Cl\,0024+16 have mid-infrared
    emission powered by star formation. At $S_{\rm 24} > 0.6$\,mJy we
    derive SFRs from the 7.7$\mu$m PAH feature of
    $\sim$30--60\,M$_\odot$\,yr$^{-1}$. Only 1/12 of our 24$\mu$m
    selected sample shows unambiguous evidence of an AGN. Our
    PAH-derived SFRs agree well with our previously estimated SFRs
    from the 24$\mu$m luminosity. In contrast to the infrared-derived
    SFRs, the typical optical ([O{\sc ii}]) derived SFRs for the same
    population are modest, with SFR([O{\sc ii}])$\lesssim
    2M_\odot$\,yr$^{-1}$. Indeed, from an optical standpoint, the
    LIRGs' spectra resemble spiral galaxies and are mainly classified
    as the potentially dusty e(a) or un-remarkable e(c) galaxies in
    the nomenclature of Dressler et al.\ (1999) and subsequent works.}

\item{The most natural fate of LIRGs in $z\sim0.5$ clusters is
    evolution into lenticular galaxies by the present day. Our
    arguments are supported by simple evolutionary models that show
    the expected luminosity evolution of the LIRGs after the cessation
    of star formation can match the bright-end of the local cluster S0
    luminosity function. At their current rates, the LIRGs in
    Cl\,0024+16 could build-up a $10^{10}$\,$M_\odot$ stellar bulge in
    $\sim 200$\,Myr. We propose that the mode of starburst in the
    LIRGs is -- like many local starbursts -- circumnuclear, although
    this has to be confirmed with higher-resolution observations
    beyond the scope of the current work. Such circumnuclear
    starbursts at $z\sim0.5$ would favour the required bulge-to-disc
    evolution of a spiral--lenticular transformation, and would
    naturally explain the presence of $\sim$4\,Gyr old stellar
    populations observed in the bulges of some local S0s. }

\end{enumerate}

Spiral galaxies falling into clusters at high redshift cannot escape
from the deep potential well: their descendants must exist in the
cores of clusters at the present day. For three decades the mystery of
the co-eval reversal in fractions of spiral/S0s in rich clusters over
the past 4\,Gyrs has not been satisfactorily explained. However, the
mid-infrared reveals the true situation: a fraction of these galaxies
are actually undergoing vigorous star formation. We argue that these
galaxies will evolve into S0s by the present day -- they could be the
missing transition galaxies that bridge the evolutionary gap between
the distant, active population of distant clusters and the passive
`red and dead' lenticulars found in the local Universe.

\section*{Acknowledgments}

The authors thank the anonymous referee for a careful and insightful
report that has improved the quality of this paper, and John Lucey,
Lauren MacArthur, Mark Norris, Russell Smith, J. D. Smith, Mark
Swinbank \& Bianca Poggianti for helpful discussions, and to Olli
Czoske for providing optical spectra for some of the LIRGs. 

JEG is supported by the U.K. Science and Technology Facilities
Council. IRS \& RSE acknowledges support from the Royal Society. TT
acknowledges support from the National Science Foundation through
CAREER award NSF--0642621, by the Sloan Foundation through a Sloan
Research Fellowship, and by the Packard Foundation through a Packard
Fellowship. This work is based on observations made with the {\it
  Spitzer Space Telescope}, which is operated by the Jet Propulsion
Laboratory (JPL), California Institute of Technology, under NASA
contract 1407. Support from NASA though a JPL grant in support of
Spitzer proposals GO\ 3143 \& GO\ 30621 is gratefully acknowledged.
 
\bibliographystyle{apj}

\end{document}

%% file: tab1.tex
\begin{table*}
\begin{center}
\caption{Properties of MIPS cluster galaxies targeted with IRS}
\scriptsize
\begin{tabular}{@{\extracolsep{\fill}}lccccccccl}
\tableline
\tableline
 & (1) & (2) & (3) & (4) & (5) & (6) & (7) & (8) \cr
Catalogue name (short name)  & R.A. & Dec. & $z$ & $S_{24}$&
$F_{7.7}$& $F_{11.3}$
&SFR(IR) & Optical class &Note \cr
& (h m s) & ($\circ$~$'$~$''$) && (mJy) & (10$^{-17}$\,W\,m$^{-2}$) &
(10$^{-17}$\,W\,m$^{-2}$) &   (M$_\odot$\,yr$^{-1}$)\cr
\tableline
MIPS~J002606.1+170416.4 (a) &        00~26~06.1 &   +17~04~16.4 &0.3904 & 1.97 & $5.6\pm0.8$ & $3.1\pm0.2$ & $34\pm10$ & --\cr
MIPS~J002721.0+165947.3 (b) &        00~27~21.0 &  +16~59~47.3  & 0.3964 & 1.48 & $9.6\pm0.6$ & $4.6\pm0.2$   & $59\pm16$ &e(a) & H$_2$S(3)?\cr
MIPS~J002621.7+171925.7 (c) &        00~26~21.7 &  +17~19~25.7  & 0.3809 &0.95 & $9.1\pm0.6$ & $2.9\pm0.2$  & $56\pm16$& e(c)  &H$_2$S(5)/[Ar~{\sc ii}]? \cr
MIPS~J002715.0+171245.6 (d) &        00~27~15.0 &  +17~12~45.6  &0.3813 & 0.95 & $5.6\pm0.9$ & $0.6\pm0.3$   & $35\pm11$ & --& Weak 11.3$\mu$m \& [Ne~{\sc ii}]?\cr
MIPS~J002652.5+171359.9 (e) &        00~26~52.5 &  +17~13~59.9  & 0.3799 &0.84 & $10.1\pm1.3$ & $3.9\pm0.3$  & $62\pm19$ & e(c) \cr
MIPS~J002609.1+171511.5 (f) &        00~26~09.1 &   +17~15~11.5 & 0.3940 & 0.77 & $4.6\pm0.9$ & $2.9\pm0.2$   & $28\pm9$ & e(c) \cr
MIPS~J002703.6+171127.9 (g)  &        00~27~03.6 &  +17~11~27.9  &0.3956 & 0.71 & $6.8\pm0.8$ & $2.3\pm0.2$   & $42\pm12$ & e(a)& [S~{\sc iv}]?\cr
MIPS~J002636.3+165926.0 (h) &       00~26~36.3 & +16~59~26.0  & 0.4063 &0.70 & v. weak  & -- & -- & k+a  & AGN (X-ray source)\cr 
MIPS~J002633.7+171221.4 (i)  &       00~26~33.7 & +17~12~21.4 & 0.3958 & 0.68 & -- & v. weak & --& e(a) & LL\,2 only \cr
MIPS~J002620.2+170407.4 (j) &       00~26~20.2 & +17~04~07.4 & 0.3963 & 0.67 & -- & weak &-- & e(a) & LL\,2 only\cr
MIPS~J002624.3+171129.8 (k)  &       00~26~24.3 & +17~11~29.8 & 0.3943&0.67 & -- & -- & -- & k & Not detected\cr
MIPS~J002637.3+170055.1 (l)  &       00~26~37.3 & +17~00~55.1  &0.3970 & 0.61 & -- & weak &-- & e(c) & LL\,2 only \cr
\tableline
\end{tabular}
\begin{minipage}{0.97\textwidth}\vspace{1mm} \footnotesize
{\sc Notes ---}(1/2) Co-ordinate is centroid of 24$\mu$m detection,
J2000. (3) Redshift determined from optical spectroscopy (Czoske et
al.\ 2001; Moran et al.\ 2007) (4) 16$''$ aperture-corrected flux
density from Geach et al.\ (2006). Typical uncertainty is 0.04\,mJy
(5/6) PAH 7.7 \& 11.3$\mu$m line flux
estimated from the spectral decomposition code {\sc pahfit} (Smith et
al.\ 2007). The total line flux is the total for the PAH line blends,
and models the individual aromatic emission modes as Drude profiles
(7) star formation rate derived from the far-infrared luminosity
(Kennicutt\ 1998), estimated from the 7.7$\mu$m line luminosity (see
\S3.4 for more details). (8) optical classification scheme from the {\it Morphs}
collaboration (Dressler et al.\ 1999) based on the equivalent widths
of [O{\sc ii}] and H$\delta$. 
\end{minipage}
\end{center}
\end{table*}